# Lattice dynamical properties of antiferromagnetic oxides calculated using self-consistent extended Hubbard functional method

Wooil Yang,[1] Bo Gyu Jang,[2] Young-Woo Son,[2] and Seung-Hoon Jhi[1*]

[1]*Department of Physics, Pohang University of Science and Technology, Pohang, 37673, Republic of Korea*

[2]*Korea Institute for Advanced Study, Seoul, 02455, Republic of Korea*

**Abstract**

We study the lattice dynamics of antiferromagnetic transition-metal oxides by using self-consistent Hubbard functionals. We calculate the ground states of the oxides with the on-site and intersite Hubbard interactions determined self-consistently within the framework of density functional theory. The on-site and intersite Hubbard terms fix the errors associated with the electron self-interaction in the local and semilocal functionals. Inclusion of the intersite Hubbard terms in addition to the on-site Hubbard terms produces accurate phonon dispersion of the transition-metal oxides. Calculated Born effective charges and high-frequency dielectric constants are in good agreement with experiment. Our study provides a computationally inexpensive and accurate set of first-principles calculations for strongly-correlated materials and related phenomena.

# 1. Introduction

Methods to calculate the properties of real materials at atomic levels have been advanced to a level of extreme accuracy [1,2]. Beyond verification and interpretation of measurement, the predictive power of the calculation enables design of materials [1–3]. Density functional theory (DFT) with approximation of electron interactions at a single-particle level like local density approximation (LDA) or generalized gradient approximation (GGA) has proven to be very accurate and computationally inexpensive [4–6]. However, some classes of materials, known as strongly-correlated materials, cannot be properly described using the one-electron-level approximation, due to delocalization error [7,8]. Transition-metal monoxides (TMOs) such as MnO, FeO, CoO, and NiO are representative strongly-correlated electronic systems. Their ground state is the insulating antiferromagnetic (AFM) type-II state and DFT fails to depict it properly [9,10]. The strong electron correlation can also affect the lattice dynamics of TMOs.

To overcome the shortcomings of DFT, the DFT+$U$ method has been proposed; it is an *ad-hoc* approach that includes the electron localization energy [9–12]. However, the on-site Hubbard $U$ term is adequate only for the Coulomb repulsion in localized sites. When nonlocal correlation becomes important [13–15], the on-site term alone is not enough to properly describe the electronic structure. Recently, the extended Hubbard method (DFT+$U$+$V$) has been proposed; it includes the Hubbard $V$ correction to consider intersite Coulomb interactions [14]. DFT+$U$+$V$ gives the band gap as accurately as the many-body $GW$ method in various materials, but with significantly reduced computational cost [14,16–18]. Also, the coexistence of on-site $U$ and intersite $V$ is effective to describe the complex materials, which have both strong electron localization and orbital hybridization [19], and to describe transition-metal dioxide molecules in which $3d$ states are overlocalized [20].

Typically, the on-site $U$ is determined empirically to match the results from measurement or many-body calculations of structural and electronic properties [9–12]. Direct calculations of the Hubbard terms require the occupation number at the atomic sites, and the nonorthogonalized atomic orbitals (NAOs) are often used as the atomic orbital projectors, to reduce computational complexity [21]. Various methods to obtain Hubbard parameters self-consistently have been proposed [22–30]. The pseudohybrid Hubbard density functional proposed by Agapito, Curtarol, and Buongiorno Nardelli (ACBN0) enables direct evaluation of the on-site $U$ by using the Hartree-Fock (HF) formalism [28,30]. Its extension to the intersite *V* is also proposed with the combination of the Löwdin orthonormalized atomic orbitals (LOAOs) [16,17]. The LOAO leads to a trace of the on-site occupation matrix compatible to the Mulliken population analysis and prevents double counting of the Hubbard correction in the overlap regions [31,32].

To address the lattice dynamical properties by using the Hubbard-corrected DFT, the forces from the Hubbard terms should be considered along with the standard DFT forces. The occupation numbers in the Hubbard terms are expressed in terms of the localized atomic orbitals, so the site-dependent Pulay forces arise naturally. The overlap matrix consists of the NAOs, so evaluation of the Pulay forces requires a very complex procedure to calculate the derivative of the square root of the overlap matrix. Recently, a computationally-efficient method to calculate the Pulay force and stress by using LOAOs has been suggested [33]. The correct Pulay forces from the Hubbard terms produce improved lattice dynamical properties of group IV semiconductors [34] and the structural and electronic properties of TMOs [32].

In this study, we used the self-consistent DFT+*U*+*V* method to perform point-by-point analysis of the structural, electronic, and lattice dynamical properties of MnO and NiO by GGA, GGA+*U*, and GGA+*U*+*V* functionals. We quantified how the lattice dynamics of strongly-correlated MnO and NiO are improved by including proper description of the Hubbard terms. The on-site *U* terms provide a primary correction for the electronic localization in transition metals, but correct computation of insulating ground states requires inclusion of the interaction between transition metal 3*d* and oxygen 2*p* states [35,36]. Quantum Monte Carlo calculations [37,38] show that the electron correlation in TMOs is gauged by the amount of hybridization between transition metal 3*d* states and oxygen 2*p* states, which is effectively considered by yielding each intersite *V* terms in the extended ACBN0 functionals.

## 2. Computational details

For all calculations, we used the QUANTUM ESPRESSO (QE) package [39] and norm-conserving pseudopotentials (NC-PP) from Pseudo Dojo library [40]. We used the GGA functional in the form of Perdew-Burke-Ernzerhof (PBE) for the electron exchange-correlation interactions [6]. We performed Brillouin-zone sampling on a $k$-point grid of 19×19×19 mesh to optimize structure and calculate electronic structure. Due to the convergence problem in the Hubbard parameters, we obtained the ground state energy in the DFT+$U$+$V$ method in three steps. First we calculated the atomic orbital occupation number in the GGA level to determine the rest of the Hubbard-corrected energy functional except the Hubbard parameters. Then we used our modified in-house QE package [16] to obtain converged Hubbard parameters self-consistently using the extended ACBN0 method. Finally, we used the converged Hubbard parameters to calculate the ground state energy of the system. The cutoff energy was set to be 110 Ry, and the self-consistency threshold for both the total energy and Hubbard parameters was $10^{-8}$ Ry. The cutoff distance for the intersite $V$ was set to include the nearest neighbors.

We used atomic orbital projectors as the local projectors to express the Hubbard-corrected energy functionals. This process enables calculation of the site-dependent Pulay force from the Hubbard-corrected energy functionals. We employed the method implemented in Ref. [33] to calculate the Pulay force and stress, as described previously [34].

To calculate the harmonic phonons, we used the PHONOPY package [41]. MnO and NiO can be regarded as having cubic rocksalt structure near room temperature. We chose experimental lattice constants of the rocksalt structure [42] and the AFM type-II order that has alternating ferromagnetic (111) planes with opposite magnetizations of cations. To calculate the interatomic force constants (IFCs), we used the frozen phonon method with a supercell of 108 atoms in the trigonal cell and the $k$-point grid of 2×2×2 mesh. We calculated the heat capacity at constant pressure within the quasi-harmonic approximation (QHA) in a supercell of 32 atoms. We calculated the ion-clamped macroscopic dielectric constants and the Born effective charges by using the finite field approach [43,44] with a $k$-point grid of 15×15×15 mesh to sample the Brillouin zone for DFT+$U$ and DFT+$U$+$V$, whereas for PBE we used density functional perturbation theory (DFPT) implemented in the QE package [39,45].

## 3. Result & Discussions

Table 1. Calculated $U$ and $V$ [eV] for 3$d$ ($U_d$) and 2$p$ states ($U_p$). $V_{dp}$ is the intersite Hubbard parameter between transition-metal 3$d$ and oxygen 2$p$ states of the nearest pair atoms.

|     | Method | $U_d$ | $U_p$ | $V_{dp}$ |
| --- | --- | --- | --- | --- |
| MnO | DFT+$U$ | 4.88 | 3.35 | |
|     | DFT+$U$+$V$ | 5.23 | 3.41 | 2.90 |
| NiO | DFT+$U$ | 8.96 | 3.91 | |
|     | DFT+$U$+$V$ | 7.65 | 2.52 | 3.15 |

We obtained converged on-site $U$ and intersite $V$ parameters for MnO and NiO by using the extended ACBN0 functionals at the experimental cubic lattice constants (Table 1). The calculated on-site $U$ parameters of NiO are larger than the values in previous reports [16,30]. These values can be varied by the choice of pseudopotentials [16,17]. With the inclusion of the intersite interactions, the on-site Hubbard parameters of 3$d$ states ($U_d$) increase in MnO but decrease in NiO. This tendency indicates that DFT+$U$ and DFT+$U$+$V$ describe the screening effects differently in these compounds.

The magnitude of $U_d$ is related to the gap between occupied and unoccupied 3$d$ bands. Therefore, the intersite $V$ terms shift the unoccupied 3$d$ bands upward in MnO and downward in NiO compared to DFT+$U$ case (figure 1). The intersite $V$ terms strengthen the hybridization between the atomic orbitals, so the localization by the on-site terms is slightly relieved [14,16].

### 3. 1. Structural and electronic properties

Table 2. Calculated and measured (Expt.) structural parameters of MnO. Lattice parameters $a$ [Å], rhombohedral angles $\theta$ [°], magnetic moments [$\mu_B$/atom] of AFM-II states, and electronic band gap $E_g$ [eV].

| Method | $a$ | $\theta$ | $\mu_B$ | $E_g$ |
|---|---|---|---|---|
| PBE | 4.4360 | 91.57 | 4.54 | 0.96 |
| DFT+$U$ | 4.5114 | 90.57 | 4.82 | 2.59 |
| DFT+$U$+$V$ | 4.4708 | 90.59 | 4.71 | 2.69 |
| Expt. | 4.4315[a] | 90.6[a] | 4.58[a], 4.79[b] | 3.9-4.1[c] |

[a]Ref. [46]

[b]Ref. [47]

[c]Ref. [48,49]

Table 3. Calculated and measured (Expt.) structural parameters of NiO. Lattice parameters $a$ [Å], rhombohedral angles $\theta$ [°], magnetic moments [$\mu_B$/atom] of AFM-II states, and electronic band gap $E_g$ [eV].

| Method | $a$ | $\theta$ | $\mu_B$ | $E_g$ |
|---|---|---|---|---|
| PBE | 4.1886 | 90.211 | 1.54 | 1.16 |
| DFT+$U$ | 4.2522 | 90.044 | 1.91 | 3.99 |
| DFT+$U$+$V$ | 4.2209 | 90.063 | 1.77 | 3.79 |
| Expt. | 4.1704[a] | 90.08[a] | 1.77[b], 1.90[a] | 4.0-4.30[c] |

[a]Ref. [46]

[b]Ref. [47]

[c]Ref. [35,48]

We calculated the structural parameters to compare the results with those by other functionals and experiments (Tables 2, 3). The experimental values were measured at low temperature ~5 K [46,47]. PBE yielded lattice parameters that were closest to experimental measurement, but overestimated the lattice distortion of the cubic symmetry, and underestimated the magnetic moment. This result confirms that PBE has delocalization error and cannot depict the ground state of TMOs correctly [9,10].

Calculations with the on-site $U$ terms alone improved the estimates of lattice distortion and magnetic moment, but still overestimated the lattice constant in both compounds. With the inclusion of the intersite $V$ terms, all values improved compared to the results with the on-site $U$ terms alone, and showed good agreement with experiment in both materials. The lattice parameter and bulk modulus of NiO by DFT+$U$+$V$ were slightly larger than or similar to the values (4.161 Å and 194 GPa, respectively) obtained using quantum Monte Carlo methods [50]. Also, the cohesive energies by DFT+$U$+$V$ (9.64 eV for MnO, 9.44 eV for NiO) were very similar to those by quantum Monte Carlo calculations (9.40 and 9.54 eV) and measurement (9.5 eV for both materials) [37,50].

We calculated band gaps of each optimized structure (Tables 2, 3). Due to the on-site $U$ of transition-metal 3$d$ states, the band gaps of MnO and NiO were closer to the measured values than those calculated using PBE. Compared to band gaps calculated using only on-site $U$ terms, inclusion of the intersite $V$ terms between transition metal 3$d$ states and oxygen 2$p$ states increased the band gap of MnO slightly but decreased the band gap of NiO slightly. This result differs from a previous calculation by DFT+$U$+$V$ method, in which the band gap increased compared to the bandgap calculated using DFT+$U$ [16]; the difference may be a result of a difference in the choice of atomic orbitals for the Hubbard corrections.

We obtained the electronic band structures and density of states (DOS) of MnO and NiO at the experimental cubic lattice parameters (Fig. 1). For MnO, the on-site $U$ correction shifted the unoccupied 3$d$ bands upward from PBE band structures and the flat occupied 3$d$ bands at about -2 eV downward from them, and thereby increased overlap between Mn 3$d$ and O 2$p$ states. In the highest valence bands of DFT+$U$, the intensity of Mn 3$d$ and O 2$p$ states are similar; this status indicates that MnO has mixed Mott-Hubbard and charge-transfer character [49]. Hybridization between Mn 3$d$ and O 2$p$ states in the occupied manifolds decreased the band width of the valence bands compared to PBE. This decrease was also observed in previous DFT+$U$ and $GW$ calculations [51,52]. When the shoulder in the experimental spectra is set to the peak of the highest valence bands of DFT+$U$, the DFT+$U$ result yields qualitative agreement with experiments from photoemission and inverse photoemission [49].

The lowest conduction band at the Γ point has mostly 4*s* character and is therefore barely influenced by the inclusion of the intersite *V* terms. The 3*d* bands in the conduction bands around 6 eV shifted further upward due to the large $U_d$ for Mn, so the bandwidth slightly increased. The overlap between Mn 3*d* and O 2*p* in the valence bands was similar between DFT+*U* and DFT+*U*+*V*, but the band width in the valence bands of DFT+*U*+*V* increased slightly.

For NiO, the on-site *U* term shifted the unoccupied 3*d* bands upward and the occupied 3*d* bands downward, compared to PBE; this is similar to the result for MnO. As a result, the parabolic 4*s* band was located below the 3*d* bands and the conduction band minimum occurred at the Γ point. Also, the top of the valence band was corrected to have O 2*p* character instead of Ni 3*d* (Fig. 1b). In DFT+*U* calculation, Ni 3*d* and O 2*p* states were clearly separated at near -4 eV from the edge of the valence bands; this result is similar to the one obtained using linear response theory [14]. With the intersite *V* term, the 3*d* bands in the conduction band lowered and the width of the conduction bands increased. The hybridization between Ni 3*d* and O 2*p* states in the valence bands increased compared to DFT+*U*. The peak from Ni 3*d* states formed near –2 eV from the edge of the valence bands, in agreement with experiments [35]. However, unlike MnO, the bandwidth in the valence bands decreased when the intersite *V* terms were considered. The occupied satellite peak from Ni 3*d* states shifted upward compared to DFT+*U* and was positioned at about –6 eV; this result is contrast to experimental results, which show a satellite structure near –8 eV from the edge of the valence bands [35]. DFT+DMFT calculations [53] indicated that to obtain agreeable spectral densities with experiments, the dynamical correlations are also important, beyond the local correlations and the hybridization between Ni 3*d* and O 2*p* states.

Compared with previous DFT+*U* calculations [52], the on-site $U_d$ is related to the gap between the unoccupied and occupied 3*d* states and the inclusion of the intersite *V* terms reduces the effect of the on-site *U* terms in the valence bands. In addition, due to the interaction between the unoccupied transition-metal 3*d* and oxygen 2*p* states, the bandwidth of 3*d* states in the conduction band was increased. Our current implementation of the intersite interactions is limited to atomic sites and azimuthal quantum numbers. To correctly consider the charge-transfer character between $3d$ and $2p$ states, the intersite $V$ terms should be expressed in full orbital dependence [14].

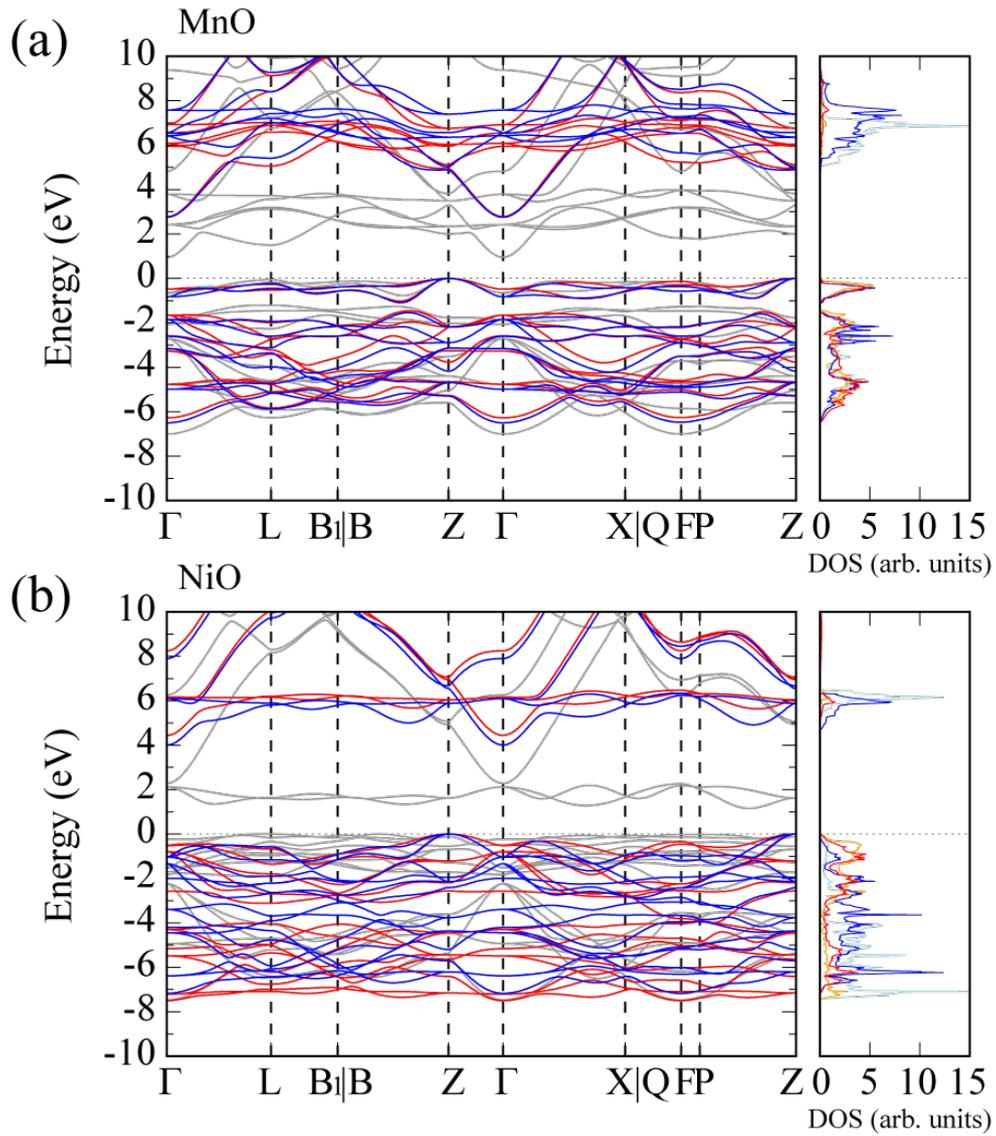

Figure 1. Left panels: Calculated electronic band structures of (a) MnO and (b) NiO along the symmetry lines in the Brillouin zone using PBE (grey), DFT+$U$ (red), DFT+$U$+$V$ (blue). Right panels: Projected density of states on the transition metal 3$d$ orbital in cyan and oxygen 2$p$ orbital in orange for DFT+$U$ calculations, and the transition metal 3$d$ orbital in blue and oxygen 2$p$ orbital in red for DFT+$U$+$V$ calculations. The top of the valence band is set at the zero energy.

Table 4. Calculated and measured (Expt.) ion-clamped macroscopic dielectric constants ($\varepsilon^\infty$), Born effective charges (Z*). DFPT (PBE) and the finite field method (DFT+$U$, DFT+$U$+$V$) are used for calculation.

|  |  | PBE | DFT+$U$ | DFT+$U$+$V$ | Expt. |
|---|---|---|---|---|---|
| MnO | Z* | 2.55 | 2.33 | 2.44 | 2.2[a] |
|  | $\varepsilon^\infty$ | 7.39 | 4.01 | 4.43 | 4.95[b] |
| NiO | Z* | 2.37 | 2.12 | 2.16 | 2.2[a] |
|  | $\varepsilon^\infty$ | 15.91 | 3.92 | 5.41 | 5.7[b] |

[a]Estimated from $\omega_{LO}$, $\omega_{TO}$ and $\varepsilon^\infty$ [54]

[b]Ref. [55]

MnO and NiO have distorted cubic symmetry in the AFM type-II state and transform to cubic rocksalt structure above the Néel temperature (MnO: 116 K; NiO: 523 K). The experimental phonon dispersions were measured at room temperature, which is above the Néel temperature of MnO, and at which the rhombohedral distortion is very small for NiO [42], so we used the experimental cubic lattice parameters to calculate the Born effective charges and the high-frequency dielectric constants (Table 4). These quantities are necessary when evaluating the dynamical matrix with the non-analytical corrections $\frac{4\pi e^2}{\Omega}\frac{[q \cdot Z_j^*]_\alpha [q \cdot Z_k^*]_\beta}{q \cdot \varepsilon_\infty \cdot q}$, where $Z_j^*$ is the Born effective charge of the $j^{th}$ atom, $\varepsilon_\infty$ is the high-frequency dielectric constant, $q$ is the wave vector, and $\alpha$ and $\beta$ are the Cartesian axes [56–58]. As a result of self-interaction error [59], PBE produced the largest (overestimated) Born effective charges for both materials. DFT+$U$ results were smaller than DFT+$U$+$V$ calculations, which are closest to the nominal charges of the TMOs. The underestimation of band gaps by PBE leads to overestimation of dielectric constants, because they are inversely proportional to the band gap. The dielectric constants by DFT+$U$+$V$ were larger than those by DFT+$U$ and comparable to the experimental results. When the experimental cubic lattice parameters were used to calculate the band gap, DFT+$U$ yielded 2.78 eV for MnO and 4.44 eV for NiO, whereas DFT+$U$+$V$ yielded 2.76 eV for MnO and 4.01 eV NiO. The lowest conduction band of MnO has 4$s$ character, so it was not affected by the Hubbard terms (Fig. 1a), as calculated band gaps showed. Still, the dielectric constants were increased substantially when the intersite $V$ term was included; this result indicates that the intersite $V$ term affects electronic screening noticeably.

Calculated Born effective charges indicate that the on-site $U$ terms alone increased the ionic character in the TMOs but the inclusion of intersite $V$ terms reduced it. The charge-density difference (Fig. 2) between PBE, DFT+$U$, and DFT+$U$+$V$ gauges the relative localization of electrons due to the Hubbard correction terms. The on-site $U$ terms drove Mott localization of electrons at atomic states so that the residual occupation of the minor spin in 3$d$ states became almost zero and the occupation of oxygen 2$p$ states was increased compared to PBE. In contrast, the intersite $V$ terms increased the number of electrons involved in Mn-O bonding, and thereby reduced the ionic character compared to DFT+$U$. We found a similar tendency in NiO except that the total occupation of the minor spin was close to 3.

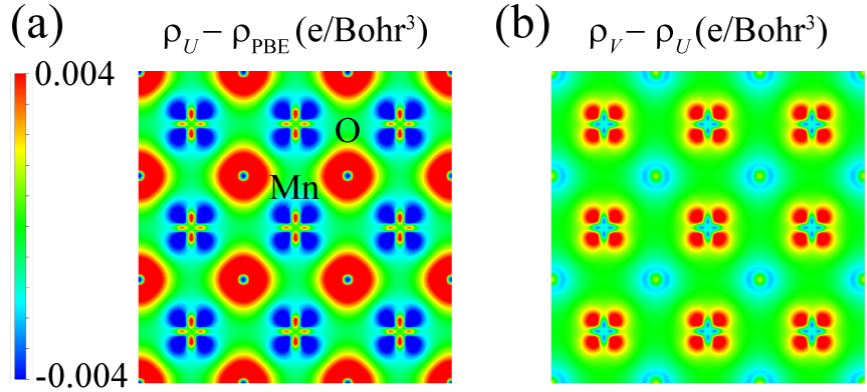

Figure 2. Plots of charge-density difference for MnO between (a) DFT+$U$ and PBE and (b) DFT+$U$+$V$ and DFT+$U$ ranging from 0.004 to –0.004 e/Bohr$^3$ to highlight regions in which charge is accumulated (red) and depleted (blue).

### 3. 2. Phonon dispersions and specific heat

We compared the lattice dynamics of MnO and NiO between PBE, DFT+$U$, and DFT+$U$+$V$, with experimental lattice constants in the cubic symmetry used to construct the primitive cell. The phonon band structures of MnO and NiO had similar dispersions overall with slight difference in the band splitting. The longitudinal optical (LO) mode at the Γ point varied depending on the Born effective charges and the high-frequency dielectric constants obtained from each functional as explained below.

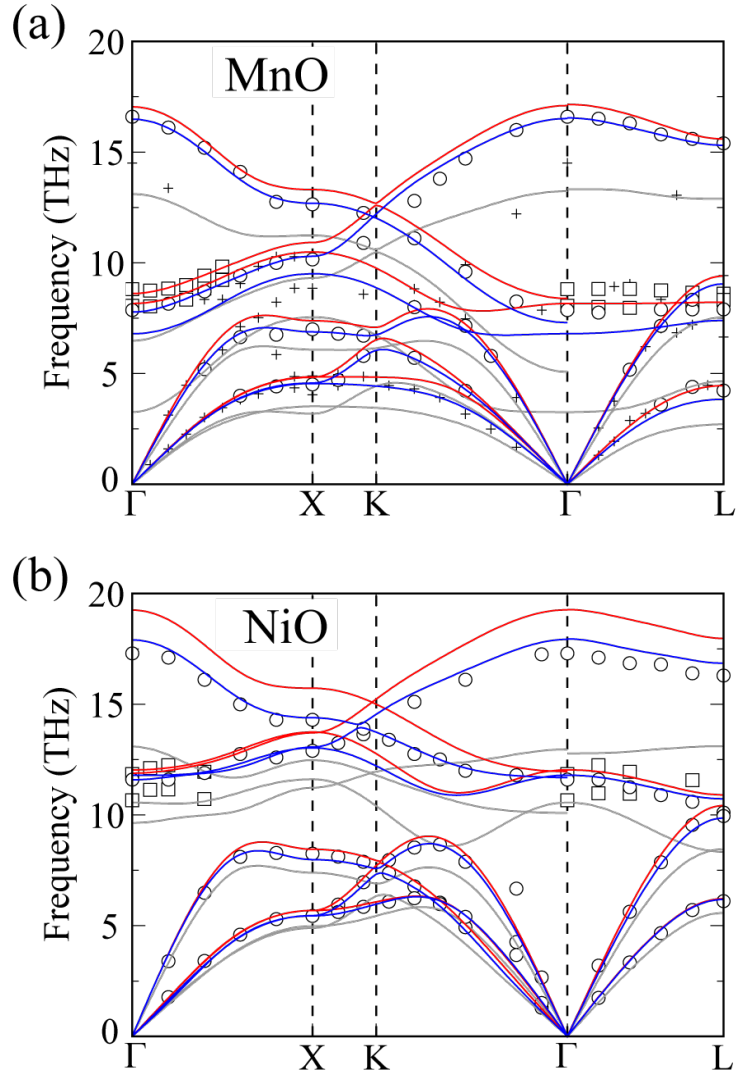

Figure 3. Calculated phonon band structures of (a) MnO and (b) NiO at the experimental lattice constants of cubic symmetry, using PBE (grey), DFT+$U$ (red), and DFT+$U$+$V$ (blue). Experimental data are included: crosses [60,61], open circles [62], and open squares [63] in (a), and open circles [64] and open squares [63] in (b).

We calculated phonon dispersions of MnO (Fig. 3a) and NiO (Fig. 3b) with PBE, DFT+$U$, and DFT+$U$+$V$ functionals along the cubic symmetry lines and together with experimental data [60–64]. The underestimation of the electronic band gap in PBE caused the calculated phonon dispersions of MnO and NiO to deviate from experiments. Especially, PBE yielded an LO frequency about 4 THz lower than experimental measurements at the Γ point.

For MnO, the inclusion of the on-site $U$ terms caused an overall upward shift of phonon dispersions and decrease in splitting between the transverse optical (TO) modes at the Γ point, compared to PBE. These changes are similar to a previous study [65].

The splitting of the TO modes along the Γ-$X$ line and the discontinuity in the TO mode along the Γ-$L$ line have a purely magnetic origin [63,65,66], because we adopted the trigonal unit cell to emulate explicitly the AFM type-II state. The TO splitting disappears in the paramagnetic phase [54]. The antiferromagnetic coupling in MnO is the super-exchange interaction mediated by nonmagnetic oxygen atoms [67], and its strength is inversely proportional to the energy separation of the unoccupied Mn 3$d$ and O 2$p$ states. The on-site $U$ terms in 3$d$ states gave an upward shift of the minority-spin $d$ states (Fig. 1), and thereby increase the energy separation and thus reduce the magnetic exchange interaction. These effects explain the reduced TO splitting when the on-site $U$ terms are considered [65].

The intersite $V$ terms favor the relative charge accumulation along Mn 3$d$ and O 2$p$ states and resultantly reduce the effect of the on-site $U$ terms. Although $U_d$ for Mn was larger in DFT+$U$+$V$ than in DFT+$U$, the intersite $V$ terms leaded to a slight increase of TO splitting compared to that obtained from DFT+$U$, and to an overall decrease in phonon frequencies compared to DFT+$U$. The phonon frequencies including non-analytic correction by calculated Born effective charges and high-frequency dielectric constants were in good agreement with measurement at room temperature [62]. Calculated phonon band structure of MnO showed that, although DFT+$U$ produces reasonably good agreement with experiment including the TO splitting along the Γ-$X$ line, the explicit consideration of the interaction between Mn 3$d$ and O 2$p$ states by the additional intersite $V$ terms gives nonnegligible correction to DFT+$U$, especially in the optical branches.

Below the Néel temperature, the rhombohedral distortion is small for NiO, and our calculated phonon dispersions of NiO in the cubic symmetry (Fig. 3b) were comparable to measurement [42]. The PBE and DFT+$U$ results exhibited similar tendency as in MnO. DFT+$U$ produced the phonon dispersion shifted upward and reduced TO splitting compared to PBE calculation. Due to the large $U_d$ of Ni, the splitting in the TO mode along the Γ-$X$ and Γ-$L$ was very small for NiO in DFT+$U$ and DFT+$U$+$V$ calculations. The splitting in the TO mode has opposite sign in NiO and MnO [65,67–69], and this effect was clearly shown in PBE calculations. The TO splitting is proportional to the derivative of the nearest neighbor magnetic coupling with respect to atomic displacement [67]. The different order in TO modes is related to the different sign of the nearest-neighbor coupling in MnO and NiO, which is explained by the Goodenough-Kanamori rule [70,71]. Due to the half-filled $d$ orbitals of Mn, the nearest-neighbor coupling for MnO is antiferromagnetic. However, for NiO, $t_{2g}$ orbitals of Ni are filled and the antiferromagnetic coupling, which is the virtual transition between $t_{2g}$ and $e_g$ orbitals of Ni atoms via the anion $p$ orbitals, is suppressed. The resultant net interaction is weakly ferromagnetic [72].

The inclusion of the intersite $V$ terms yielded the phonon dispersions of NiO that matched with measurement [64], and thereby alleviated the overestimations, especially for LO modes. In DFT+$U$ calculation, the LO modes were greatly overestimated mostly as a consequence of the non-analytic correction. In DFT+$U$, the charge localization by the on-site $U$ terms reduced the Coulomb screening, as seen in the high-frequency dielectric constants (Table 4), and leaded to an excessive upward shift of LO mode. The inclusion of the intersite $V$ terms recovered the dielectric screening properly to produce overall phonon frequencies that matched well with measurements. However, in NiO, $U_d$ decreased when the intersite interaction was included (Table 1). In DFT+$U$+$V$ calculations, this change caused a more drastic change of LO mode in NiO than in MnO.

The improvement of the phonon band structures, particularly the optical modes, by the DFT+$U$+$V$ method was attributed to the correct estimate of the Born effective charge and the band gap. The improvement is also a result of correct estimation of the magnetic state, because the bond strength and thus the lattice vibrations are properly produced when the occupancy of the major and minor spin states is correct. The monomorphous nonmagnetic states are unstable and hypothetical [73], so fully converged phonon dispersions were obtained at the antiferromagnetic state.

Using calculated phonon dispersions, we studied the heat capacity at constant pressure within the QHA in 2×2×2 supercell of 32 atoms. We checked the size dependency of the Helmholtz free energy to find the difference between 2×2×2 and 3×3×3 supercell results < 10 meV/unit cell for all three functionals.

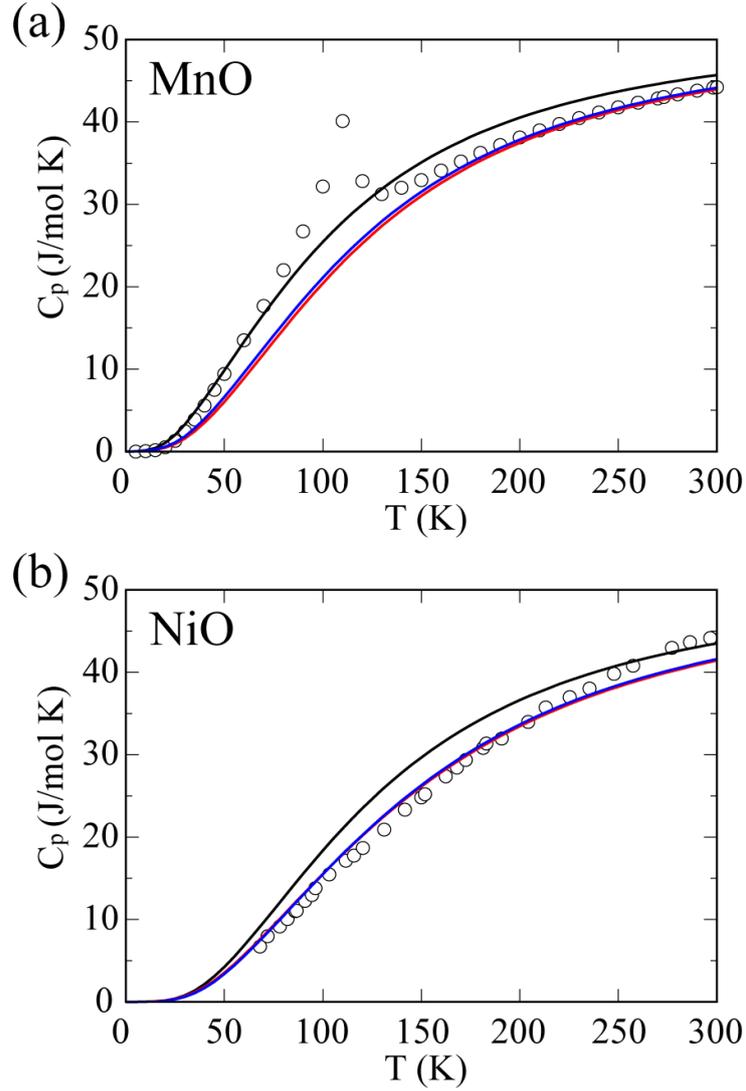

Figure 4. Calculated lattice contribution to the specific heat capacity at constant pressure in (a) MnO and (b) NiO. PBE (black), DFT+$U$ (red), DFT+$U$+$V$ (blue) under the quasi-harmonic approximation. Open circles: measurements (Ref. [74] for MnO, [75] for NiO).

We calculated lattice contribution to the heat capacity of MnO (Fig. 4a) and NiO (Fig. 4b) together with measurement. DFT+$U$ and DFT+$U$+$V$ produced very similar results and significant improvement from PBE. Although the phonon dispersions by DFT+$U$ and DFT+$U$+$V$ differed in the optical branches, they were similar in the acoustic branches; this comparison explained the almost-identical heat capacity estimated by DFT+$U$ and DFT+$U$+$V$ at low temperature. DFT+$U$+$V$ calculation showed good agreement with measurement, but showed some deviation, especially near the Néel temperature. A sharp peak at about 120 K for MnO in measurement indicates the magnetic transition [74,76]. We did not consider the magnetic degrees of freedom in our calculations, so we expect that calculated values are smaller than measurements, especially below the Néel temperature. The magnetic state was primarily depicted by the on-site $U$ term [73], but the intersite $V$ was also required for accurate description of the magnetic state (Table 3). For NiO, slight overestimation of DFT+$U$+$V$ at low temperature may be due to the choice of the cubic structure rather than the distorted rhombohedral structure.

## 4. Conclusion

We studied the lattice dynamical properties of antiferromagnetic transition-metal oxide MnO and NiO by using the extended Hubbard functional that incorporates the on-site $U$ and intersite $V$ terms fully self-consistently into the DFT method. These self-consistent on-site $U$ and intersite $V$ terms are truly necessary to properly describe the ground states and dynamical properties of the TMOs. The structural parameters and the electronic structures calculated with the Hubbard terms have improved accuracy significantly by the mean-field type calculations and show improved agreement with experiment. Especially, the dielectric constants, which are the measure of the Coulomb screening, were sensitively influenced by the intersite interactions. Calculated lattice dynamical properties of MnO and NiO were also correctly described with the inclusion of the intersite $V$ terms. For the specific heat capacity, the calculations by DFT+$U$+$V$ showed excellent agreement with the measurements above the Néel temperature, but the magnetic contribution should be included for appropriate comparison at low temperature. Our work demonstrates that the fully self-consistent DFT+$U$+$V$ method can provide a consistent and parameter-free framework to explore the electronic properties, lattice-dynamical properties, and electron-phonon interactions in strongly-correlated materials, without a need for significant computational resources.

Acknowledgement

This paper was supported by the National Research Foundation of Korea (NRF; Grant No. 2018R1A5A6075964) funded by the Korean government (MSIT). Y.-W.S. was supported by the National Research Foundation of Korea (NRF) (Grant No. 2017R1A5A1014862, SRC program: vdWMRC center), KIAS individual Grant No. (CG031509) and acknowledged the computational services from CAC of KIAS.